# Tuning Magnetism in Ising-type van der Waals Magnet FePS$_3$ by Lithium Intercalation

Dinesh Upreti[1], Rabindra Basnet[1,2], M M Sharma[1], Santosh Karki Chhetri[1], Gokul Acharya[1], Md Rafique Un Nabi[1,3], Josh Sakon[4], Bo Da[5], Mansour Mortazavi[3], Jin Hu[1,3*]

[1]Department of Physics, University of Arkansas, Fayetteville, AR 72701, USA

[2]Department of Chemistry & Physics, University of Arkansas at Pine Bluff, Pine Bluff, Arkansas 71603, USA

[3]MonArk NSF Quantum Foundry, University of Arkansas, Fayetteville, Arkansas 72701, USA

[4]Department of Chemistry & Biochemistry, University of Arkansas, Fayetteville, AR 72701, USA

[5]Center for Basic Research on Materials, National Institute for Materials Science, Tsukuba, Ibaraki 305-0044, Japan

Abstract

Recently, layered materials transition metal thiophosphate $M$P$X_3$ ($M$ = transition metals, $X$ = S or Se) have gained significant attention because of their rich magnetic, optical, and electronic properties. Specifically, the diverse magnetic structures and the robustness of magnetism in the two-dimensional limit have made them prominent candidates to study two-dimensional magnetism. Numerous efforts such as substitutions and interlayer intercalations have been made to tune the properties of these materials, which has greatly deepened the understanding of the underlying mechanisms that govern the properties. In this work, we focus on modifying the magnetism of Ising-type antiferromagnet FePS$_3$ using electrochemical lithium intercalation. Our work unveils the effectiveness of electrochemical intercalation as a controllable tool to modulating magnetism, including tuning magnetic ordering temperature and inducing low temperature spin-glass state, offering an approach for implementing this material into applications.

*jinhu@uark.edu

I. Introduction

Emerging magnetism in van der Waals (vdWs) magnets provides a deeper understanding of phenomena arising from two-dimensional (2D) magnetism such as magnetic skyrmions[1,2] and other spin textures[3], which shed light on future spintronic applications[4–7]. This has attracted significant efforts to effectively manipulate magnetic ordering and spin orientations in vdW magnetic materials, such as chemical substitution [8–10], high-pressure [11–14], and strain effect [15]. The magnetic vdW-type transition metal thiophosphate $M$P$X_3$ ($M$ = transition metals, $X$ = S or Se) is a large antiferromagnetic (AFM) material family, in which each transition metal $M$ carries localized moments in a honeycomb lattice and sandwiched by chalcogen $X$ atom layers[16–19]. $M$P$X_3$ compounds exhibit AFM ground states in bulk form. Such long-range magnetic orders persist even in the two-dimensional (2D) limit for certain $M$P$X_3$ such as FePS$_3$ and MnPS$_3$ [20–22]. The magnetic properties of $M$P$X_3$ are strongly dependent on the choice of $M$ and $X$, which has motivated extensive metal $M$[23–32] and chalcogen $X$[8,33,34] substitution studies to tune magnetism. In addition, one effective route to tune magnetic properties in vdW materials is inter-layer intercalation of guest ions, atoms, or molecules[35–45]. The layered structure of $M$P$X_3$ compounds allows for inter-layer intercalation of guest species such as pyridine, lithium (Li), and ammonia which has also been demonstrated as an efficient approach to tune magnetism in $MnPS_3$, $NiPS_3$, and $FePS_3$[36–38,40,42–44]. These intercalation studies have revealed signatures of ferrimagnetism[37,40], weak ferromagnetism[36], and spin glass[38], as well as the modulation of magnetic ordering temperatures[36,42–44].

Among various $MPX_3$ compounds, $FePS_3$ displays Ising-type magnetism that is characterized by out-of-plane magnetic moments[46]. The Ising-type magnetism in $FePS_3$ has been proposed to stem from large crystal electric field arising from the strong trigonal distortion of the $FeS_6$ octahedra[47,48]. Because of the strong magnetic anisotropy associated with the Ising-type magnetism, the long-range magnetic order in $FePS_3$ persists down to the single-layer[20]. However, given the zero net magnetization of the AFM ground state, the magnetic characterization in atomically thin layers is challenging. So far, the persistence of the 2D AFM order of $FePS_3$ has been probed using Raman spectroscopy which is an indirect probe[20]. This has hindered the potential application of $FePS_3$ though it is one of the first known 2D magnets. Establishing a ferromagnetic (FM) order in $FePS_3$ would make the study of 2D magnetism more accessible and convenient by direct magnetic characterization tools such as Kerr rotation [5,49] and magnetic force microscope[50]. Unfortunately, magnetism in $FePS_3$ is rather robust against common tuning approaches such as chemical substitution in either metal or chalcogen sites[33]. Earlier theoretical studies[51,52] have predicted that charge doping may induce a FM order in $MPX_3$. Hence, compared to isovalent substitutions, inter-layer intercalation that usually results in charge doping might be an effective strategy to induce ferromagnetism in $MPX_3$. Previous intercalation studies on $FePS_3$ using molecular intercalants have revealed a possible spin-glass state[38] and a reduction in magnetic ordering temperature [43]. In addition, lithium is another good intercalant because of its smaller atomic size which may allow for a greater amount of intercalant without strong structure modifications. Li intercalation on FePS$_3$ has been studied a few decades ago but both unchanged [53] and suppressed[44] transition temperature have been reported, which motivates us to revisit the effect of Li intercalation on magnetism in FePS$_3$.

In this study, we performed systematic electrochemical intercalation of lithium in FePS$_3$ and studied the evolution of magnetic properties. From the magnetic susceptibility measurements, the AFM transition temperature is found to reduce for heavily intercalated samples. In addition, the zero-field cooling (ZFC) and field cooling (FC) susceptibility exhibit substantial irreversibility and magnetic hysteresis near zero magnetic fields at low temperatures, which are likely attributed to weak ferromagnetism due to spin-glass state or vacancy-induced magnetism. Such engineering of magnetism in FePS$_3$ offers a promising platform to tune Ising-type anisotropy that could be extended to other Ising-type magnetic systems.

## II. Experiment

Single crystals of FePS$_3$ were synthesized using a chemical vapor transport method with I$_2$ as the transport agent. The stoichiometric mixture of Fe, P, and S powder was vacuum-sealed in a quartz tube and placed in a two-zone furnace with a temperature gradient from $750^0$-$650^0$C for one week. The electrochemical intercalations of Li into FePS$_3$ single crystal were performed in an electrochemical workstation (Battery testing system 8.0, Neware) in a similar way reported previously[37]. The elemental composition and crystal structure of the obtained FePS$_3$ crystals were determined by energy dispersive x-ray spectroscopy (EDXS) and x-ray diffraction (XRD), respectively. The magnetic properties were measured using a Magnetic Property Measurement System (MPMS3, Quantum Design).

Different techniques such as liquid/wet chemical intercalation[36,38,39,54,55], gaseous intercalation[56,57], and electrochemical intercalation[37,53,58] has been adopted to intercalate layered compounds to tune their properties. The liquid chemical process is carried out by immersing the host in a guest solution. However, this process is easily accompanied by the exfoliation of the bulk materials to layers which hinders the further applications[59]. Additionally, the degree of intercalation is not easily controllable. Gaseous intercalations are performed by a vapor transport technique[56,57] in which the host material and guest intercalant are placed separately at cold and hot zones respectively in a two-zone furnace. The gaseous form of the intercalant is transferred to the cold zone and intercalates into the host material. While this method has several benefits such as scalability and a higher degree of intercalation, it is irreversible which inhibits the potential application in batteries, sensors, and optical switches[59]. Compared to the above methods, electrochemical intercalation is a controllable approach[59,60] for which the amount of intercalant can be controlled by the applied electrical current and the duration of intercalation[37]. Moreover, the intercalation is reversible, making it suitable for devices and energy storage applications[59].

Figures 1(a) and 1(b) show the schematic of the electrochemical intercalation of Li in the vdW gap of $FePS_3$ single crystals. In this setup, the $FePS_3$ single crystals and Li-chips act as anode and cathode, respectively, where both electrodes were dipped in an electrolytic solution of Li bis-trifluoro methane sulfonamide salt in dimethoxy-ethane and dioxolane mixed in the ratio of 1:1. The intercalation was performed using a coin battery setup as reported previously[37], and the batteries were prepared inside an argon-filled glovebox. Electrochemical intercalation was carried out with a constant discharge current of 20 $\mu$A similar to the previous Li-intercalation study on $NiPS_3$[37]. To promote homogeneous intercalation, the experiments were performed at 50º C. The amount of Li intercalation was controlled by varying the intercalation time. $FePS_3$ single crystals with similar mass were chosen to ensure controllable study. Because of the difficulty of probing Li using the common x-ray-based techniques such as XRD and EDS, it is not possible to accurately determine the content of the intercalated Li. Instead, we label the samples as sample #1 (pristine $FePS_3$), #2, #3, #4, and #5, in which the Li amount systematically increases according to the duration of the electrochemical intercalation process and lattice constant, as will be shown below.

III.     **Result and Discussion**

The images of the sample #1 to #4 are shown in Fig. 1(c). Compared with the pristine sample, crystals somewhat lose metal luster upon increasing intercalation time. For sample #5 with the longest electrochemical intercalation and hence the highest amount of Li, it became powder-like and we did not include it in comparison. A similar trend was observed for Li intercalated $NiPS_3$[37]. The crystal structures of the intercalated samples were characterized by powder x-ray diffraction (PXRD). As shown in Figure 1c, sample #2 and #3 display highly similar diffraction patterns as the pristine sample (sample #1) except for some tiny peak shift, indicating that the crystal structure of $FePS_3$ is maintained up on Li intercalation. For sample #4, main diffraction peaks are broadened but still match with that of the pristine sample. The emergence of additional diffraction peaks as indicated by asterisks is likely due to the formation of $Li_3PS_4$ or $Li_4P_2S_6$ on the edge of the sample[37,61]. The intercalated Li atoms are expected to occupy the vdW gap, but the exact location cannot be identified by common X-ray-based techniques. Nevertheless, the evolution of crystal lattice with intercalation provides clear evidence for successfully intercalation. As shown in Fig. 1c, Rietveld structure refinement reveals an unchanged space group of $C2/m$ up to sample #4. For sample #5, the refinement can only yield lattice parameters without reliable atomic positions. Therefore, in this work, we limit the studies to sample #1 to #4. Table 1 summarizes the refined lattice parameters. Overall, the in-plane lattice parameters *a* and *b* show subtle variations up on Li intercalation, which has also been observed in the pyridine-intercalated MnPS$_3$[36]. In contrast, the out-of-plane lattice parameter *c* shows minor changes for samples #2 and #3 but increases strongly by 0.83% for sample #4 and 9.7% for #5. The above observations clearly reveal the control of Li amount by the duration of electrochemical intercalation, and suggest that the doped Li most likely intercalates into the vdW gap, as illustrated in Fig. 1b.

Previous studies have revealed that intercalation on *M*P*X*$_3$ successfully tunes magnetic properties including the emergence of ferrimagnetism[37,40], ferromagnetism[36], and spin glass[38], as well as the tuning of ordering temperature [36,42,43]. To study the effects of Li intercalation on FePS$_3$, the temperature and field dependences of magnetization were examined. The pristine FePS$_3$ is characterized by Ising-type antiferromagnetism with out-of-plane moment orientation below 120 K [47,62] . The Ising-type magnetism in FePS$_3$ is manifest by substantial anisotropy even in the paramagnetic (PM) state, with the out-of-plane magnetic susceptibility $\chi_\perp$ (measured with an out-of-plane magnetic field *H*⊥*ab*) much larger than the in-plane one $\chi_{//}$ (measured with an in-plane magnetic field *H*//*ab*)[47]. As shown in Fig. 2, the strong anisotropy with much greater $\chi_\perp$ in the PM state is also observed in our Li-intercalated FePS$_3$ samples, implying robust Ising-type magnetism against intercalation in FePS$_3$. Compared to the higher temperature PM state, the low-temperature AFM state is more significantly affected by Li intercalation. The pristine FePS$_3$ displays nearly temperature-independent susceptibility in the AFM state (Fig. 2a), whereas strong low-temperature susceptibility upturns emerge in Li-intercalated samples (Figs. 2b-2d). Furthermore, unlike the perfectly overlapped ZFC and FC susceptibility in FePS$_3$, Li-intercalated samples display clear irreversibility at low temperatures, as indicated by $T_{\text{irr}}$ (black triangles) in Figs. 2b-2d. These observations suggest a possible spin-glass state or vacancy-induced ferromagnetism up on Li intercalation, as will be discussed later.

In addition to the modification to the low-temperature susceptibility, Li intercalation also tunes the AFM ordering temperature $T_N$ in FePS$_3$. As shown in Fig. 2a, at the temperature that magnetic susceptibility reaches maximum ($T_{\text{max}} \approx 130$K, dashed lines), FePS$_3$ displays a broad susceptibility hump instead of a sharp transition. AFM transition characterized by sharp susceptibility drop occurs at lower temperature of 120 K (pink dotted lines). This susceptibility hump above $T_N$ has also been seen in previous studies on FePS$_3$ and many other $M$P$X_3$ compounds and has been attributed to the in-plane short-range magnetic correlations of the layered $M$P$X_3$[47]. Increasing the Li content from sample #2 to sample #4 reduces both $T_{\text{max}}$ and $T_N$, as shown in Figs 2c-2d. The susceptibility hump above $T_N$ is gradually suppressed and the temperature difference between $T_{\text{max}}$ and $T_N$ is reduced. The observed evolution of $T_{\text{max}}$ and $T_N$ with Li intercalation is summarized in Fig. 3. Because of the difficulty in determining the exact content of Li as stated above, here we use the interlayer spacing $d$ which measures the vdW gap to characterize the degree of Li intercalation, because intercalating Li elongates $d$ as revealed by XRD discussed above. As shown in Fig. 3, systematic suppression of $T_N$ with intercalation from 120 K in the pristine FePS$_3$ to 91 K in sample #4 is seen, which is accompanied by an elongation of $d$ by 0.8%. Our direct observation of suppression of $T_N$ in FePS$_3$ by intercalation is consistent with the previous report in which magnetic transition is determined by Raman scattering[44]. Moreover, unlike the need for more than 50% increase in $d$ to suppress $T_N$ by 35%[43], in this work 24% $T_N$ reduction is realized by less than 1% change in $d$.

To obtain some insights into the correlation between interlayer distance and $T_N$, in Table 2 we summarize the relative changes of interlayer distance and $T_N$ normalized to that of the pristine samples for FePS$_3$[38,43], together with MnPS$_3$ [36,40] and NiPS$_3$[37,43] which have been relatively intensively studied by intercalation. Both Li- and organic-ion-intercalations are included in the table, which facilitates comparison of a wide range of changes in $d$. Interestingly, magnetic ordering temperature in FePS$_3$ and MnPS$_3$ can be suppressed by intercalation, but it is much more robust in NiPS$_3$ against 55% increase in $d$. Such a difference might be associated with the different mechanisms for magnetism in those materials. For example, it has been reported that, because magnetism is governed by direct exchange in MnPS$_3$ but superexchange in NiPS$_3$, $T_N$ is only slightly suppressed in MnPS$_3$ but strongly enhanced in NiPS$_3$ by Se substitution[24] . For the intercalation study, considering the unchanged insulating nature and the accompanied $d$ enhancement, the observations summarized in Table 2 suggest that the interlayer magnetic exchange $J_z$ might play a non-trivial role in governing magnetism in FePS$_3$ and MnPS$_3$, which is generally overlooked since magnetism in $M$P$X_3$ family has been generally considered to be mainly governed by the intralayer exchange interactions [62,63]. Indeed, recent studies suggest that interlayer exchange interaction may not be negligible in $M$P$X_3$ [64,65]. Nevertheless, this appears conflict to the experimental observation of the very robust AFM long-range order with unchanged $T_N$ in the FePS3 monolayer[20].

In addition to the role of interlayer interaction, other mechanisms might also give rise to reduced $T_N$. An early Raman study on Li-intercalated FePS$_3$[44] has attributed the $T_N$ reduction to the suppression of in-plane exchange interactions caused by magnetic fluctuations from a random distribution of lithium. In addition, the lowering of magnetic ordering temperature is also observed in the pyridine-intercalated MnPS$_3$[36], which has been ascribed to magnetic dilution due to the vacancies. Though PXRD is unable to reveal vacancies, their presence might be supported by our magnetic susceptibility measurements. As stated above, FePS$_3$ displays a broad hump at susceptibility maxima above $T_N$ similar to many other $M$P$X_3$ compounds (Fig. 2). The temperature $T_{max}$ for such hump is relatively more efficiently suppressed and eventually disappears (i.e., merge with $T_N$) by Li intercalation. Since such susceptibility hump is believed to be caused by the in-plane short-range magnetic correlation above $T_N$[47], its suppression can be attributed to the suppression of in-plane exchange. Furthermore, the $T_N$ reduction may also be associated with carrier doping which has been observed in heavily doped europium chalcogenides[66]. However, this is inconsistent with the fact that the Li-intercalated FePS$_3$ samples remain highly insulating. To elucidate the actual mechanism for $T_N$ suppression, additional theoretical and experimental effort is needed.

In addition to the diminution of magnetic ordering temperature, a bifurcation between ZFC and FC susceptibility has been observed at a temperature below $T_N$, at around 20 K for samples #2 and #3, and 66 K for sample #4, as denoted by $T_{irr}$ in Fig 2 and summarized in Fig. 3. Such irreversibility implies the rise of the FM interactions at low temperatures, which is further supported by the field dependence of magnetization. As shown in Fig. 3, up on Li intercalation, magnetic hysteresis loop gradually develops in out-of-plane magnetization (measured with $H\perp ab$) at $T = 2$ K but is negligible in in-plane magnetization (measured with $H//ab$). Both out-of-plane and in-plane magnetizations lack saturation behavior up to 7 T, suggesting that the ferromagnetic correlations, if exist, may not develop into a long-range order. Such ferromagnetic correlations might be caused by magnetic impurities, vacancy magnetism, or moment canting. The presence of magnetic impurities is less likely because no such impurities have been probed by XRD as described above. Vacancies in the metal sites may occur, which has also been proposed in the 4-aminopyridine intercalated FePS$_3$[45]. In addition, vacancies may also give rise to moment canting. Careful crystal structure and magnetic structure studies are needed to clarify such a scenario.

In addition to ferromagnetic correlations, the development of a spin glass state also explains the observed irreversibility. Spin glass can arise with strong geometric or magnetic frustrations. Given the comparable FM nearest-neighbor interaction $J_1$ and AFM third nearest-neighbor interaction $J_3$ in FePS$_3$[46], the magnetic frustration arising from competing FM $J_1$ and AFM $J_3$ is possible when intercalated Li disturbs the exchange interactions. Indeed, spin glass states upon intercalation in A[NH$_3$]$_x$FePS$_3$ (A = Li, K) [38] and metal substitution in Mn$_{1-x}$Fe$_x$PS$_3$[32] have been reported. Interestingly, magnetic irreversibility has not been observed in Li-intercalated NiPS$_3$[37], which can be understood in terms of the lack of strongly competing magnetic exchange interactions since the magnetism in NiPS$_3$ is dominated by the much stronger AFM $J_3$ interaction[63].

## IV. Conclusion

In conclusion, we have successfully intercalated lithium in the FePS$_3$ single crystals using an electrochemical approach. The Ising type magnetism in the pristine FePS$_3$ is found to be robust towards lithium intercalation, while a suppression of $T_N$ occurs with high amount of Li amount. The reduction in $T_N$ suggests that the interlayer interactions may play a role in stabilizing magnetic ordering temperature in some $M$P$X_3$ compounds. In addition to the decrease in $T_N$, magnetic irreversibility below $T_N$ is seen along with the magnetic hysteresis, which might be attributed to a spin glass state arising from the magnetic frustrations caused by intercalation. The successful tuning of magnetic ordering temperature and spin glass state with Ising-type magnetism intact in FePS$_3$ offers an alternative and efficient path in modulating magnetism of 2D materials. The controllable electrochemical intercalation and its easy integration in devices further provides an approach for implementing this material in applications.

**Acknowledgments**

This work was primarily (crystal growth and intercalation) supported by the U.S. Department of Energy, Office of Science, Basic Energy Sciences program under Grant No. DE-SC0022006. M.R.U.N acknowledges the MonArk NSF Quantum Foundry, which is supported by the National Science Foundation Q-AMASE-i program under NSF award No. DMR-1906383. R.B, M.M.S, and M.M acknowledges µ-ATOMS, an Energy Frontier Research Center funded by DOE, Office of Science, Basic Energy Sciences, under Award No. DE-SC0023412 (Refinement and part of the magnetic property analysis). J. S. acknowledges the support from NIH under award P20GM103429 for the powder XRD experiment.

# Figures

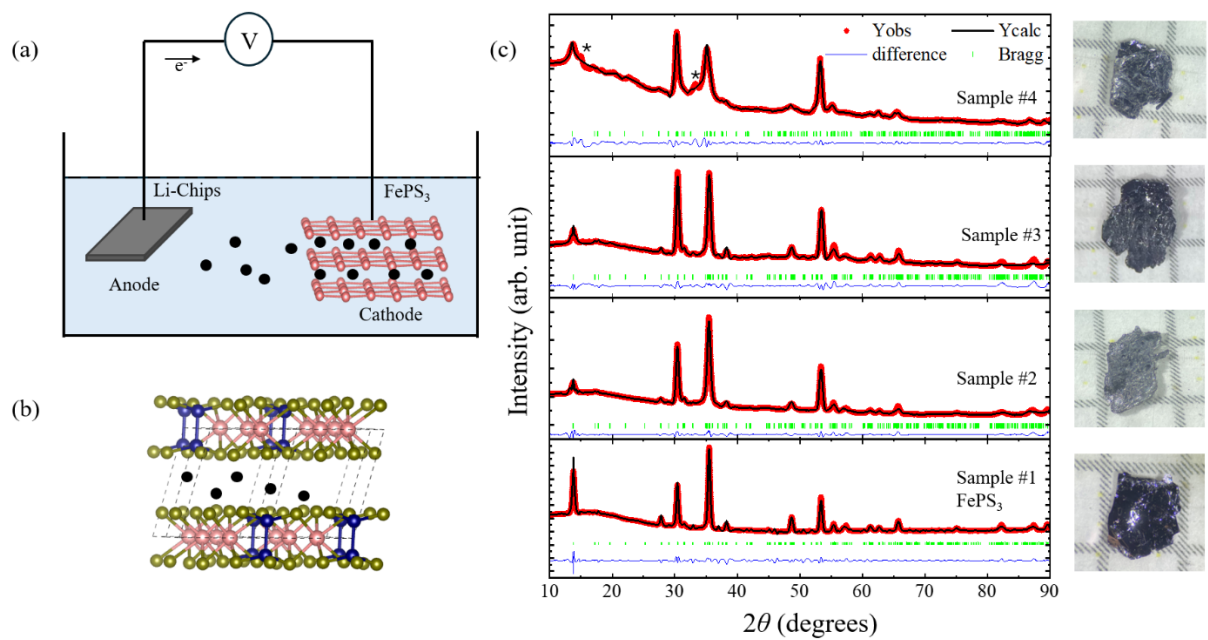

FIG. 1. (a) Conceptual schematic of the electrochemical intercalation. The intercalation was performed using a battery setup, but the concept is the same. (b) Schematic of the Li intercalation into the vdW gap of FePS$_3$. (c) Powder x-ray diffraction patterns and Rietveld refinement for crystal structures of pristine FePS$_3$ (sample #1) and various Li-intercalated (samples #2 to #4) FePS$_3$. Images of FePS$_3$ and intercalated crystals are shown in the right panel.

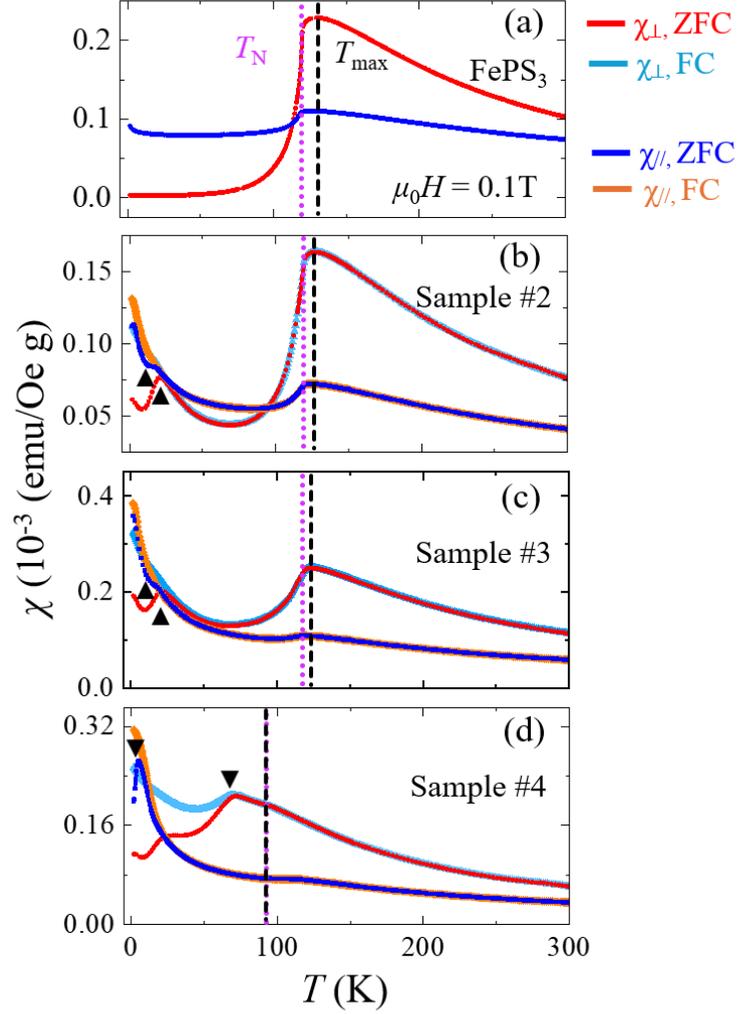

FIG. 2. Temperature dependence of out-of-plane ($\chi_\parallel$) and in-plane ($\chi_\perp$) susceptibilities ($\chi$) for (a) the pristine FePS$_3$ (samples #1) and (b-d) Li-intercalated FePS$_3$ (samples #2-4). Both ZFC and FC susceptibilities are shown in intercalated samples. For the pristine FePS$_3$ in (a), only ZFC data are shown because its ZFC and FC data overlap. The vertical pink dotted lines and the black dashed lines denote $T_N$ and $T_{max}$, respectively. The black triangles denote $T_{irr}$, below which irreversibility is seen.

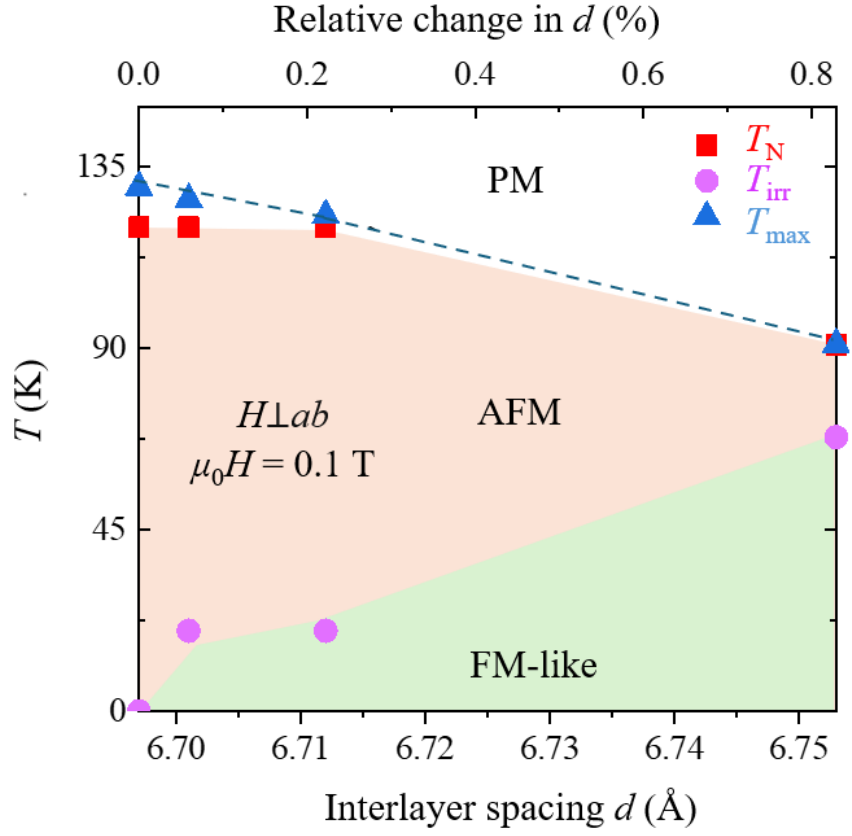

FIG. 3. Magnetic phase diagram for Li-intercalated FePS$_3$, showing the evolution of magnetic phases with temperature and interlayer spacing $d$. The phase boundaries are determined based on the out-of-plane susceptibility ($\chi_{//}$) measured under 0.1 T magnetic field shown in Fig. 2.

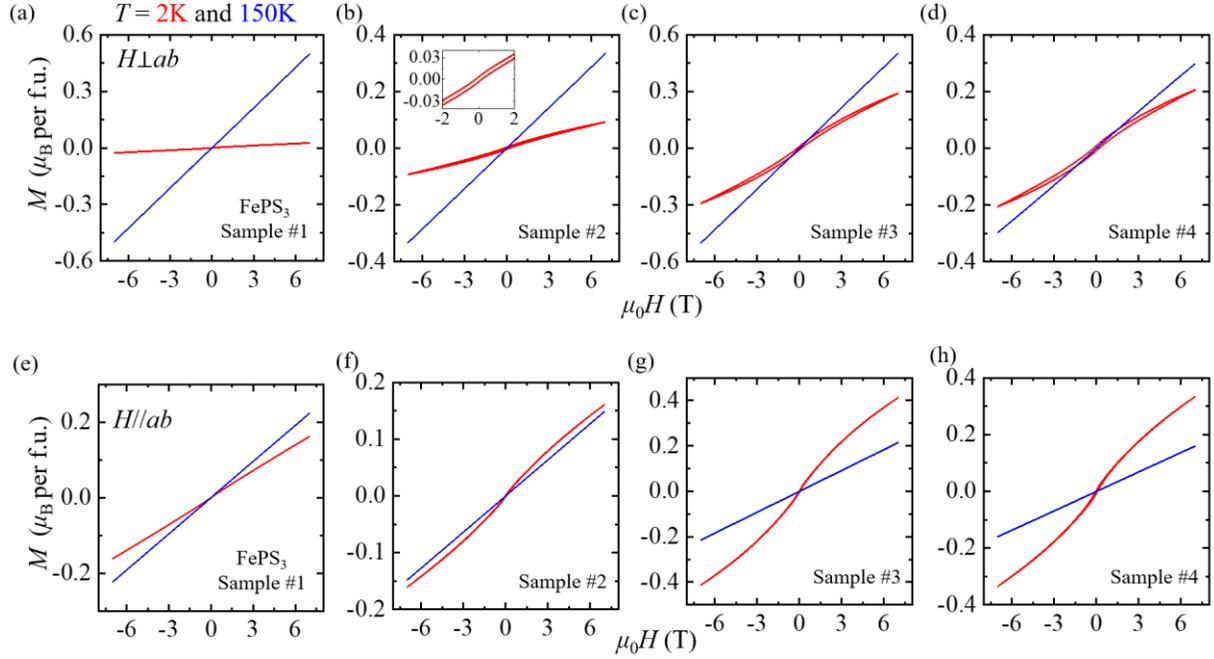

FIG. 4. Field-dependence of magnetization of the pristine (sample #1) and Li-intercalated (samples #2 to #4) FePS$_3$ at $T = 2$ K (red lines) and 150 K (blue lines), measured with out-of-plane ($H \perp ab$, panels a-d) and in-plane ($H // ab$, panels e-h) magnetic fields. Inset in (b) shows the low field hysteresis loop.

**Tables**

**Table 1**. Lattice parameters *a*, *b*, and *c* obtained from Rietveld refinement for the pristine and Li-intercalated $FePS_3$.

| Sample | *a* | *b* | *c* |
|---|---|---|---|
| #1 ($FePS_3$) | 5.94353 | 10.29546 | 6.69756 |
| #2 | 5.93050 | 10.29169 | 6.70115 |
| #3 | 5.95282 | 10.27902 | 6.71267 |
| #4 | 5.95840 | 10.31935 | 6.75311 |
| #5 | 5.94500 | 10.30390 | 7.35000 |

**Table 2**. Relative changes of magnetic ordering temperatures and inter-layer distances $d$ in FePS$_3$, MnPS$_3$, and NiPS$_3$ up on intercalation.

| Host materials | Relative change in $d$ | Type of intercalation | Relative change in ordering temperature | Reference |
| --- | --- | --- | --- | --- |
| **FePS$_3$** | 54.7% | EMIM, electrochemical | -35% | 43 |
| **FePS$_3$** | 47.3% | Li, liquid ammonia | Not reported | 38 |
| **FePS$_3$** | 0.8% | Li, electrochemical | -24.17% | This work |
| **MnPS$_3$**[*] | 162.6% | Amino acid, ion exchange reaction | -35.90% | 40 |
| **MnPS$_3$**[#] | 91.31% | Pyridine, liquid | -23.07% | 36 |
| **NiPS$_3$** | 55.72% | EMIM, electrochemical | -2.02% | 43 |
| **NiPS$_3$** | 0.357% | Li, electrochemical | -1.28% | 37 |

[*]Ferrimagnetism

[#]Weak ferromagnetism but show negative Weiss temperature